\documentclass[prl,showpacs,showkeys,preprint,double-spaced]{revtex4}
\usepackage{amssymb}
\usepackage{eurosym}
\usepackage{graphicx}
\usepackage{bm}
\usepackage{tabularx}
\usepackage{multirow}
\usepackage{url}
\usepackage{euscript}
\usepackage{epsfig}
\usepackage{rotating}

\begin{document}

\title{Magnon Condensation and Spin Superfluidity}
\author{Yury~M.~Bunkov$^{a}$ and Vladimir L. Safonov$^{b,c}$}
\affiliation{$^{(a)}$ Kazan Federal University, Kremlevskaya 18, 420008 Kazan, Russia\\
$^{(b)}$ Mag and Bio Dynamics, Inc., Granbury, TX 76049, USA\\
$^{(c)}$ Physical Science Department, Tarrant County College - South Campus,
Fort Worth, TX 76119, USA}

\begin{abstract}
We consider the Bose-Einstein condensation (BEC) of quasi-equilibrium
magnons which leads to spin superfluidity, the coherent quantum transfer of
magnetization in magnetic material. The critical conditions for excited
magnon density in ferro- and antiferromagnets, bulk and thin films, are
estimated and discussed. It was demonstrated that only the highly populated
region of the spectrum is responsible for the emergence of any BEC. This
finding substantially simplifies the BEC theoretical analysis and is surely
to be used for simulations. It is shown that the conditions of magnon BEC in
the perpendicular magnetized YIG thin film is fulfillied at small angle,
when signals are treated as excited spin waves. We also predict that the
magnon BEC should occur in the antiferromagnetic hematite at room
temperature at much lower excited magnon density compared to that of
ferromagnetic YIG. Bogoliubov's theory of Bose-Einstein condensate is
generalized to the case of multi-particle interactions. The six-magnon
repulsive interaction may be responsible for the BEC stability in ferro-
and antiferromagnets where the four-magnon interaction is attractive.
\end{abstract}

\maketitle


\section{Introduction}

Spin deviations from the magnetic order in a magnetic material (ferromagnet,
antiferromagnet or ferrites) are manifested by spin waves and their quanta,
magnons. Magnons are quasiparticles which represent a very useful quantum
theoretical tool to describe various dynamic and thermodynamic processes in
magnets in terms of magnon gas. Since magnons have magnetic moments, the
external alternating magnetic field can excite extra magnons and increase
the disorder in the magnetic system. However, in certain conditions, the
increase of magnon density leads to a new state, so-called, magnon
condensate, in which a macroscopic number of magnons forms a coherent
quantum state (see, e.g., \cite{review5}, \cite{safonov}). This macroscopic
state can significantly change the properties of magnon gas, its dynamics
and transport. An example is the phenomenon of quasi-equilibrium
Bose-Einstein condensation (BEC) of excited magnons on the bottom of their
spectrum as a single-particle long-range coherent state of quantum liquid. This state generate an
uniform long-lived precession of spins formed by quantum specificity of
the magnon gas when the magnon density exceeds certain critical value. The
spatial gradients of this state exhibit a spin superfluidity, the non-potential
transport of deflected magnetization. The spin superfluidity is an extremely
interesting phenomenon for both fundamental and applied studies. 
It should be emphasized that the main paradigm
of magnetic dynamics, the Landau-Lifshitz-Gilbert equation, does not contain complete information
about the Bose-Einstein condensate of magnons.
BEC is the principal result of quantum statistics and for magnons it can exist at room
and even higher temperatures.

For the first time the existence of quasi-equilibrium Bose condensate was
demonstrated in the experiment with nuclear magnons in the superfluid
antiferromagnetic liquid crystal $^3$He-B in 1984 \cite{HPD}. The
theoretical explanation of this phenomenon \cite{Fomin} was developed on the
basis of global Ginzburg-Landau energy potential. A similar approach was
later developed to explain the atomic BEC \cite{PitaevskiiStringari2003}. In
the experiments with an antiferromagnetic $^3$He-B, the following phenomena
were observed: a) transport of magnetization by spin supercurrent between
two cells with magnon BEC; b) phase-slip processes at the critical current;
c) spin current Josephson effect; d) spin current vortex formation; d)
Goldstone modes of magnon BEC oscillations. Comprehensive reviews of these
studies can be found in Refs.\cite{volovik,BunkovVolovik2008,Bunkov2009}.
Currently magnon BEC found in different magnetic systems: i) in
antiferromagnetic superfluid $^3$He-A \cite{Sato,HPDA}; ii) in in-plane
magnetized yttrium iron garnet Y$_{3}$Fe$_{5}$O$_{12}$ (YIG) film (with two
minima in the magnon spectrum) \cite{YIGBEC1,YIGBEC2} and in normally magnetized YIG film \cite{YIGBEC3}; iii) in
antiferromagnets MnCO$_{3}$ and CsMnF$_{3}$ with Suhl-Nakamura indirect
nuclear spin-spin interaction \cite{UFN,AFM1,BBKT}.
An explanation of analogy between the atomic and magnon BEC
is given in Ref. \cite{Bunkov}. 

A microscopic theory of quasi-equilibrium magnon BEC was developed in Refs.%
\cite{KSpump}-\cite{KSjmmm} ("KS theory"). It was predicted that the
external strong pumping of magnons leads to a rapid growth of magnon density
and saturation. This state can be considered in terms of weakly non-ideal
gas of "dressed" magnons in a thermodynamic quasi-equilibrium with an
effective chemical potential $\mu $ and effective temperature $T$. The
dressed magnon energy is defined by $\varepsilon _{\mathbf{k}}=\varepsilon _{%
\mathbf{k}}^{(0)}+\delta \varepsilon _{\mathbf{k}}$, where $\varepsilon _{%
\mathbf{k}}^{(0)}=\hbar \omega _{\mathbf{k}}$ is the energy spectrum of bare
magnons and $\delta \varepsilon _{\mathbf{k}}$ is the energy shift due to
magnon gas nonlinearities. Magnon-magnon scattering processes retain the
total number of dressed magnons in the system and hold their distribution
function of the form
\begin{equation}
n_{\mathbf{k}}=\left( \exp {\frac{\varepsilon _{\mathbf{k}}-\mu }{k_{B}T}}%
-1\right) ^{-1}.  \label{Bose dist}
\end{equation}%
The instability at $\mu =\min \varepsilon _{\mathbf{k}}$ in the
quasi-equilibrium magnetic system is an analog of BEC phenomenon for the
bottom dressed magnons. The distribution (\ref{Bose dist}) seems to underlie
the phenomenon of spin superfluidity, since it nullifies the integral of
four-magnon collisions

\begin{eqnarray}
I^{(4)}\{n_{\mathbf{k}}\} &\propto &\int
d^{3}k_{1}d^{3}k_{2}d^{3}k_{3}\left\vert \Phi _{4}(\mathbf{k},\mathbf{k}_{1};%
\mathbf{k}_{2},\mathbf{k}_{3})\right\vert ^{2}  \label{INT st} \\
&&\times \left[ (n_{\mathbf{k}}+1)(n_{\mathbf{k}_{1}}+1)n_{\mathbf{k}_{3}}n_{%
\mathbf{k}_{4}}-n_{\mathbf{k}}n_{\mathbf{k}_{1}}(n_{\mathbf{k}_{2}}+1)(n_{%
\mathbf{k}_{3}}+1)\right]  \nonumber \\
&&\times \delta (\varepsilon _{\mathbf{k}}+\varepsilon _{\mathbf{k}%
_{1}}-\varepsilon _{\mathbf{k}_{2}}-\varepsilon _{\mathbf{k}_{3}})\Delta (%
\mathbf{k}+\mathbf{k}_{1}-\mathbf{k}_{2}-\mathbf{k}_{3})  \nonumber
\end{eqnarray}%
and thus this energy loss channel vanishes.

KS theory qualitatively explained the parallel pumping experiments (\cite%
{kin neust},\cite{melkov} (YIG at room temperature) and \cite{govorkov}
(nuclear magnons in CsMnF$_{3}$), where the accumulation of magnons at the
bottom of the spin wave spectrum was observed. One and a half decade later,
purposeful experiment \cite{YIGBEC1} directly demonstrated BEC of quasi
equilibrium magnons in the thin film of YIG. Subsequent experimental studies
have shown qualitative correspondence with the predicted distribution of
excited magnons \cite{disrib func} and agreement with the BEC under noisy
pumping \cite{noise}.

In this paper we analyze critical conditions of quasi-equilibrium magnon BEC
in ferro- and antiferromagnets, bulk and thin films, and evaluate the
possibilities of their experimental achievements.

\section{BEC of Bose particles}

Let us first briefly discuss BEC of real bose particles. Their distribution
is defined by Eq.(\ref{Bose dist}), where $\varepsilon _{k}={(\hbar k)^{2}}/{%
2m}$ is the kinetic energy of particle with the wave vector $k$ and mass $m$%
. The total number of bosons in the system is
\begin{equation}
N(\mu ,T)=N=V_{s}\int n_{\mathbf{k}}\frac{d^{3}k}{(2\pi )^{3}},
\label{Ntotal}
\end{equation}%
where $V_{s}$ is the volume of the system. For the critical condition $\mu
=\min \varepsilon _{k}$, from (\ref{Ntotal}) follows well-known formula for
the BEC critical temperature versus the density of bosons:
\begin{equation}
T_{BEC}=\kappa _{0}\,\frac{\hbar ^{2}}{k_{B}m}\left( \frac{N}{V_{s}}\right)
^{2/3},\text{ \ \ \ }\kappa _{0}=\frac{2\pi }{\left[ \zeta \left( \frac{3}{2}%
\right) \right] ^{2/3}}\simeq 3.31.  \label{TBECpart}
\end{equation}%
It is interesting to note that the BEC is formed mainly by bosons with high
populations when Eq.(\ref{Bose dist}) can be written as
\begin{equation}
n_{k}\simeq \frac{k_{B}T}{\varepsilon _{k}-\mu }.  \label{HTP}
\end{equation}%
Let us prove it by direct calculation. Substituting the high temperature
population (\ref{HTP}) into Eq.(\ref{Ntotal}) and cutting the upper integral
limit by the thermal energy $\varepsilon _{T}\simeq k_{B}T$, one obtains:
\begin{equation}
T_{BEC}\simeq \tilde{\kappa}_{0}\,\frac{\hbar ^{2}}{k_{B}m}\left( \frac{N}{%
V_{s}}\right) ^{2/3},\ \text{\ \ \ }\tilde{\kappa}_{0}=\frac{\pi ^{4/3}}{%
2^{1/3}}\simeq 3.65.  \label{TBECpartApprox}
\end{equation}%
We see that the only difference between Eqs.(\ref{TBECpart}) and (\ref%
{TBECpartApprox}) is a slightly different ($\sim 10\%$) numerical factor.

The fact that the high population Eq.(\ref{HTP}) is dominant does not mean
that the BEC phenomenon is a classical one. The criterion of classical
Maxwell-Boltzmann statistics $\exp \left( \mu /k_{B}T\right) $ $\ll 1$ in
this case can be written as (see, e.g., \cite{Kerzon Huang}):

\begin{equation}
\exp \left( \mu /k_{B}T\right) =\left[ \frac{V_{s}}{N}\int \exp \left( -{%
\frac{\varepsilon _{\mathbf{k}}}{k_{B}T}}\right) \frac{d^{3}k}{(2\pi )^{3}}%
\right] ^{-1}\ll 1,  \label{MB gas}
\end{equation}%
or

\begin{equation}
\frac{N}{V_{s}}\, \lambda^3_T =
\frac{N}{V_{s}}\left( \frac{2\pi \hbar ^{2}}{mk_{B}T}\right) ^{3/2}\ll 1,
\label{ClassStat}
\end{equation}%
where $\lambda_T$ is the thermal de Broglie wavelength.
Substituting BEC temperature Eq.(\ref{TBECpartApprox}) into (\ref{ClassStat}%
), we obtain the opposite relation: $2.26>1$, which obviously corresponds to
a degenerate bose gas.

\section{BEC of magnons}

Now let us consider a Bose-Einstein condensation of so-called, ``dressed"
magnons as an instability in the externally pumped quasi-equilibrium magnon
gas. The total number of magnons $N(\mu,T)$ is equal to the number of
thermal magnons $N(0,T)$ at a given temperature $T$ and the number of
magnons $N_{p}$ created by external pumping. So far as the energy shift of
dressed magnons is usually much less than the energy of bare magnons $\delta
\varepsilon_{k} \ll \min \varepsilon^{(0)} _{k} $, for simplicity we can
approximate $\varepsilon_{k} \simeq \varepsilon^{(0)}_{k} $.

\subsection{BEC in a ferromagnet}

Consider a ferromagnet with the quadratic spectrum (we neglect details of
the dipole-dipole interactions):
\begin{equation}
\varepsilon _{k}=\varepsilon _{0}+\varepsilon _{ex}\left( ak\right) ^{2}.
\label{F spectrum}
\end{equation}
Here $\varepsilon_{ex}$ is the exchange interaction constant and $a$ is the
elementary cell linear size. The quasi-equilibrium BEC will be mainly
determined by pumping if the number of pumped magnons is much greater than
the thermal magnon number $N_{p}\gg N(0,T)$. In this case we obtain an
analog of Eq.(\ref{TBECpart}):

\begin{equation}
T_{BEC}=\kappa _{0}\,\frac{2\varepsilon _{ex}}{k_{B}}\left( a^{3}\frac{N_{p}}{%
V_{s}}\right) ^{2/3},  \label{TBECquasipart}
\end{equation}%
or,
\begin{equation}
T_{BEC}\simeq \tilde{\kappa}_{0}\,\frac{2\varepsilon _{ex}}{k_{B}}\left( a^{3}%
\frac{N_{p}}{V_{s}}\right) ^{2/3}  \label{TBECquasipartApprox}
\end{equation}%
in the high-population approximation.

The above formula, however, does not work for a BEC estimate if $%
N_{p}\lesssim N(0,T)$. Using a high-population approximation, we write
\begin{eqnarray}
N_{p} &=&N(\mu ,T)-N(0,T)  \nonumber \\
&\simeq &V_{s}\int_{\varepsilon _{0}}^{\varepsilon _{T}}\left( \frac{k_{B}T}{%
\varepsilon _{k}-\mu }-\frac{k_{B}T}{\varepsilon _{k}}\right) \frac{k^{2}dk}{%
2\pi ^{2}},  
\label{TBECpump1}
\end{eqnarray}%
and obtain at $\mu =\varepsilon _{0}$
\begin{eqnarray}
\frac{N_{p}}{V_{s}} &\simeq &\frac{k_{B}T_{BEC}}{4\pi a^{3}}\,\frac{%
\varepsilon _{0}^{1/2}}{\varepsilon _{ex}^{3/2}},  \nonumber \\
T_{BEC} &\simeq &4\pi \,\frac{\varepsilon _{ex}}{k_{B}}\left( \frac{%
\varepsilon _{ex}}{\varepsilon _{0}}\right) ^{1/2}\left( a^{3}\frac{N_{p}}{%
V_{s}}\right) .  \label{NpumpBEC}
\end{eqnarray}%
These formulas coincide with the accuracy of notations with the exact
calculation given in Ref.\cite{safonov}. This is one more direct proof that
BEC is formed by the high-populated part of spectrum.

An estimate for YIG, where $\varepsilon _{ex}a^{2}/\hbar =0.092$ cm$^{2}$s$%
^{-1}$ for\ $\varepsilon _{0}/\hbar =2\pi \times 2.5$ GHz gives $%
T_{BEC}\simeq 2.14\times 10^{-17}(N_{p}/V_{s})$ cm$^{3}$K. Thus, we obtain a
room-temperature BEC $T_{BEC}\simeq 300$ K at the pumped magnon density $%
N_{p}/V_{s}=1.41\times 10^{19}$ cm$^{-3}$ that in order of magnitude
corresponds to the experiment \cite{YIGBEC1}.

As in the above case of particles, the opposition of high density of magnons
to their high population makes the classical criterion $\exp \left[ \left(
\mu -\varepsilon _{0}\right) /k_{B}T\right] \ll 1$ inapplicable to the case
of condensation when $\mu =\varepsilon _{0}$. As in the previous section, we
can write the criterion for Maxwell-Boltzmann statistics

\begin{equation}
\exp \left[ \left( \mu -\varepsilon _{0}\right) /k_{B}T\right] =\left[ \frac{%
V_{s}}{N(\mu ,T)}\int \exp \left( -{\frac{\varepsilon _{ex}\left( ak\right)
^{2}}{k_{B}T}}\right) \frac{d^{3}k}{(2\pi )^{3}}\right] ^{-1}\ll 1,
\label{magnon MB}
\end{equation}%
or

\begin{equation}
\frac{N(\mu ,T)}{V_{s}}\, \lambda^3_T = 
\frac{N(\mu ,T)}{V_{s}}\left( \frac{4\pi \varepsilon _{ex}a^{2}}{k_{B}T}%
\right) ^{3/2}\ll 1.  \label{magnon criterion}
\end{equation}%
In the case of high temperatures we have $N(\mu ,T)\simeq N(0,T)$.
Substituting the thermal density

\begin{equation}
\frac{N(0,T)}{V_{s}}=\left( \frac{k_{B}T}{2\kappa _{0}\varepsilon _{ex}a^{2}}%
\right) ^{3/2}  \label{thermal magnons}
\end{equation}%
into (\ref{magnon criterion}), we obtain the opposite relation: $2.61>1$. In
other words, magnon gas which undergoes Bose-Einstein condensation is always
a degenerate bose gas. Thus the assertion of recent publication \cite{Kopec}
``that the experimentally observed condensation of magnons in yttrium-iron
garnet at room temperature is a purely classical phenomenon" is untenable.

\subsection{BEC in an antiferromagnet}

Consider now the magnon energy of the form
\begin{equation}
\varepsilon_{k}=\sqrt{\varepsilon _{0}^{2}+\varepsilon _{ex}^{2}(ak)^{2}}.
\label{AF spectrum}
\end{equation}%
This is typical for magnons in the "easy-plane" (or, canted)
antiferromagnets. Taking into account that%
\[
k=\frac{\sqrt{\varepsilon _{k}^{2}-\varepsilon _{0}^{2}}}{a\varepsilon _{ex}}%
\text{ \ \ \ \ and \ \ \ }kdk=\frac{\varepsilon _{k}d\varepsilon _{k}}{%
\varepsilon _{ex}^{2}a^{2}},
\]%
in the high-population approximation, one can write
\begin{equation}
\frac{N(\mu =\varepsilon _{0},T)}{V_{s}}\simeq \frac{k_{B}T}{2\pi ^{2} }%
\frac{1}{a^{3}\varepsilon_{ex}^{3}}\int_{\varepsilon _{0}}^{\varepsilon _{T}}%
\sqrt{\frac{\varepsilon +\varepsilon _{0}}{\varepsilon -\varepsilon _{0}}}%
\varepsilon d\varepsilon .  \label{AF N muT}
\end{equation}%
If $N_{p}\gg N(0,T)$, for $k_{B}T\gg \varepsilon _{0}$ we obtain
\begin{equation}
T_{BEC}\simeq (2\pi )^{2/3}\,\frac{\varepsilon _{ex}}{k_{B}}\left( a^{3}\frac{%
N_{p}}{V_{s}}\right) ^{1/3}.  \label{BEC AF Np}
\end{equation}

If $N_{p}\lesssim N(0,T)$, one can rewrite Eq.(\ref{TBECpump1}) in the form:%
\begin{equation}
\frac{N_{p}}{V_{s}}\simeq \frac{k_{B}T}{2\pi ^{2}}\frac{\varepsilon _{0}}{%
a^{3}\varepsilon _{ex}^{3}}\int_{\varepsilon _{0}}^{\varepsilon _{T}}\sqrt{%
\frac{\varepsilon +\varepsilon _{0}}{\varepsilon -\varepsilon _{0}}}%
d\varepsilon .  \label{density BEC AFEP}
\end{equation}%
After integration, at $k_{B}T\gg \varepsilon _{0}$ we obtain
\[
\frac{N_{p}}{V_{s}}\simeq \frac{(k_{B}T_{BEC})^{2}}{2\pi ^{2}}\frac{%
\varepsilon _{0}}{a^{3}\varepsilon _{ex}^{3}},
\]%
or,
\begin{equation}
T_{BEC}\simeq \sqrt{2}\pi \,\frac{\varepsilon _{ex}}{k_{B}}\left( \frac{%
\varepsilon _{ex}}{\varepsilon _{o}}\right) ^{1/2}\left( a^{3}\frac{N_{p}}{%
V_{s}}\right) ^{1/2}.  \label{TBECaf}
\end{equation}%
Note that the BEC temperature for antiferromagnet has lower power dependence
on small parameter $a^{3}N_{p}/V_{s}\ll 1$ and therefore one can expect much
lower densities of pumped magnons to achieve condensation. An estimate for $%
\alpha -$Fe$_{2}$O$_{3}$ (hematite), where $\varepsilon _{ex}a/\hbar \approx
24\times 10^{5}$ cm$/$s for \ $\varepsilon _{0}/\hbar =2\pi \times 2.5$ GHz
gives \ $T_{BEC}\approx 10^{-6}(N_{p}/V_{s})^{1/2}$ cm$^{3/2}$K. Thus we
obtain a room-temperature BEC,\ $T=300$ K at$\ $ $N_{p}/V_{s}=0.89\times
10^{17}$ cm$^{-3}$.
This estimate is more lower than
corresponding estimate for a ferromagnetic YIG. This means that hematite is
a very attractive object to observe BEC of magnons experimentally.

\section{BEC in a ferromagnetic film}

Let us now consider an ultra-thin ferromagnetic film. There are two
principal cases: 1) external magnetic field $\mathbf{H}$ is parallel to the
the film surface and 2) $\mathbf{H}$ is perpendicular to this surface. In
the first case, the BEC condition $\mu =\min \hbar \omega _{\mathbf{k}}$
gives us two minima at $\pm \mathbf{k}_{\min }\neq 0$. This case was
demonstrated experimentally for the YIG film in Ref.\cite{YIGBEC1}, where
the critical density of pumped magnons was estimated numerically. Later, in
Ref.\cite{review5} it was considered analytically. Here we focus on the
second case with just one energy minimum at $\mathbf{k}=0$.

The magnon spectrum of the perpendicular magnetized ultra-thin ferromagnetic
film can be written as \cite{kakazei}:
\begin{equation}
\omega _{k}=\left\{[\omega _{H}+\omega _{ex}(ak)^{2}][\omega _{H}+\omega
_{ex}(ak)^{2}+\omega _{M}f(k\tau )]\right\}^{1/2},  \label{1}
\end{equation}%
where $\gamma =2\pi \;2.8$ MHz/Oe is the gyromagnetic ratio, $\omega
_{H}=\gamma H_{i}$, $H_{i}=H_{e}-4\pi M_{s}+H_{\perp }$ is an effective
internal magnetic field, $H_{\perp }$ is the perpendicular anisotropy field,
$M_{s}=139$ Oe is the saturation magnetization, $\omega _{M}=4\pi \gamma
M_{s}=2\pi \times 4.9$ GHz, $\omega _{ex}a^{2}=2\pi \times 1.09\times
10^{-2} $ Hz cm$^{2}$ is the exchange constant, $f(k\tau )=1-[1-\exp {%
(-k\tau )}]/k\tau $, $\tau $ is the thickness of the film. All numerical
parameters are given for YIG. Spin waves are assumed to be propagated only
in the film plane and there is a uniform magnetization along the depth.

The critical density of pumped magnons at temperature $T$ can be estimated
by the following equation:
\begin{equation}
\frac{N_{p}}{A_{s}}=\frac{N(\mu ,T)}{A_{s}}-\frac{N(0,T)}{A_{s}},  \label{3}
\end{equation}%
where the sample volume is replaced by the film area $A_{s}$. The Eq.(\ref{3}%
) in the high-population approximation has the form:
\begin{equation}
\frac{N_{p}}{A_{s}}\approx \frac{k_{B}T}{2\pi }\int_{0}^{k_{T}}\left( \frac{1%
}{\hbar \omega _{k}-\mu }-\frac{1}{\hbar \omega _{k}}\right) kdk,  \label{6}
\end{equation}
where $k_{T}$ corresponds to the frequency $\omega_{k} \simeq k_{B} T/\hbar$%
. This integral at $\mu \rightarrow \hbar \omega _{0}$ has a logarithmic
divergence for an infinitely large film. However, for a finite film we have
a magnetostatic mode on the bottom, which can be separated from the
spin-wave spectrum by a small gap $\Delta \omega $. Thus, we can estimate
Eq.(\ref{6}) as
\begin{equation}
\frac{N_{p}}{A_{s}}\approx \frac{k_{B}T}{4\pi \hbar \omega _{ex}a^{2}}\ln {%
\left( \frac{\omega _{0}}{\Delta \omega }\right) }.  \label{8}
\end{equation}%
Note that this formula corresponds to the model of ultra-thin film, in which the magnon dynamics is considered in two dimensions.
Magnetic excitations across the film plane are discrete and they interact weakly with magnons in the plane of the film. 
For this reason they practically do not affect the quasi-equilibrium in the two-dimensional system.
Formula (\ref{8}) is convenient for simple estimates, which can be refined by numerical calculations with
accounting for all magnetic degrees of freedom.

\subsection{Critical angle}

Let us now find a critical angle of the magnetic moment deviation from the
equilibrium, which is assumed to correspond to the critical number of
excited magnons. This angle is defined by the ratio of perpendicular spin
component to its longitudinal component $\tan \theta ={S_{\perp }}/{S_{z}} $%
. The perpendicular component is equal to
\begin{eqnarray}
S_{\perp } &=&\sqrt{S_{x}^{2}+S_{y}^{2}}=\sqrt{\frac{S_{+}S_{-}+S_{-}S_{+}}{2%
}} \\
&\approx &\sqrt{2Sa^{\ast }a}=\sqrt{2SN_{p}}.  \nonumber  \label{10}
\end{eqnarray}%
Substituting $S_{z}\simeq S$, for small angles one obtains
\begin{equation}
\theta \approx \sqrt{\frac{2N_{p}}{S}}=\sqrt{\frac{2\hbar \gamma }{M_{s}}%
\frac{N_{p}}{A_{s}\tau }}.  \label{11}
\end{equation}
For the film thickness $\tau =1 $$\mu $m, $\omega _{0}=2\pi \times 2.5$ GHz,
$\Delta \omega =2\pi \times 1$ Hz at room temperature $T=300$ K we have $%
\theta _{film}\approx 0.044$ ($2.5^{\circ}$). An estimate for a bulk material at the same
conditions gives $\theta _{bulk}\approx 0.061$ ($3.5^{\circ}$). Taking into account finite
thickness of the film, one can expect the experimental value of the angle
will be within $\theta _{film}<\theta <\theta _{bulk}$. For comparison, the
BEC deflection angle in the antiferromagnetic liquid crystal $^3$He-B was
about $10^{-3}$ \cite{review5}.

\section{BEC stability}

The critical density is required but not a sufficient condition for a
uniform magnon BEC. For the BEC stability it is important to consider the
interaction between magnons. The Hamiltonian of magnon system described by
creation ($b_{\mathbf{k}}^{\dagger }$) and annihilation ($b_{\mathbf{k}}$)
bose operators can be written in the form:
\begin{equation}
\mathcal{H}=\sum\limits_{\mathbf{k}}(\varepsilon _{\mathbf{k}}-\mu )b_{%
\mathbf{k}}^{\dagger }b_{\mathbf{k}}+\mathcal{H}_{4}+\mathcal{H}_{6}+%
\mathcal{H}_{8}+...  \label{Hamiltonian}
\end{equation}%
where the magnon-magnon interaction terms are
\[
\mathcal{H}_{4}=\frac{1}{2}\sum_{\mathbf{1,2;3,4}}\Phi _{4}(\mathbf{k}_{1},%
\mathbf{k}_{2};\mathbf{k}_{3},\mathbf{k}_{4})b_{\mathbf{1}}^{\dagger }b_{%
\mathbf{2}}^{\dagger }b_{\mathbf{3}}b_{\mathbf{4}}\Delta (\mathbf{k}_{1}+%
\mathbf{k}_{2}-\mathbf{k}_{3}-\mathbf{k}_{4}),
\]
\begin{eqnarray*}
\mathcal{H}_{6} &=&\frac{1}{3}\sum_{\mathbf{1,2,3;4,5,6}}\Phi _{6}(\mathbf{k}%
_{1},\mathbf{k}_{2},\mathbf{k}_{3};\mathbf{k}_{4},\mathbf{k}_{5},\mathbf{k}%
_{6})b_{\mathbf{1}}^{\dagger }b_{\mathbf{2}}^{\dagger }b_{\mathbf{3}%
}^{\dagger }b_{\mathbf{4}}b_{\mathbf{5}}b_{\mathbf{6}} \\
&&\times \Delta (\mathbf{k}_{1}+\mathbf{k}_{2}+\mathbf{k}_{3}-\mathbf{k}_{4}-%
\mathbf{k}_{5}-\mathbf{k}_{6})
\end{eqnarray*}%
and so fourth. According to Bogoliubov's theory (see, e.g.,\cite{safonov},%
\cite{BogoBogo}), we have to single out classical condensate amplitudes with
$\mathbf{k=0}$: $b_{\mathbf{0}}^{\dagger }=b_{\mathbf{0}}=\sqrt{N_{0}}$, $%
N_{0}=N-N^{\prime }$. Here $N=\sum\nolimits_{\mathbf{k}}b_{\mathbf{k}%
}^{\dagger }b_{\mathbf{k}}$ is the total magnon number and $N_{0}$ is the
magnon number in the condensate. Assuming that $N^{\prime }/N\ll 1$, we can
reduce the Hamiltonian (\ref{Hamiltonian}) to $\mathcal{H}=\mathcal{H}_{0}+%
\mathcal{H}_{2}$, where
\begin{equation}
\mathcal{H}_{0}=(\varepsilon _{\mathbf{0}}-\mu )N_{\mathbf{0}}+\frac{1}{2}%
\mathcal{T}_{4}(0)N_{0}^{2}+\frac{1}{3}\mathcal{T}_{60}N_{0}^{3}+...
\label{zerothH}
\end{equation}%
is the condensate energy and
\begin{equation}
\mathcal{H}_{2}=\sum\limits_{\mathbf{k\neq 0}}\left[ A_{\mathbf{k}}b_{%
\mathbf{k}}^{\dagger }b_{\mathbf{k}}+\frac{B_{\mathbf{k}}}{2}\left( b_{%
\mathbf{k}}b_{-\mathbf{k}}+b_{\mathbf{k}}^{\dagger }b_{-\mathbf{k}}^{\dagger
}\right) \right] .  \label{QuadForm}
\end{equation}%
Here
\begin{eqnarray}
A_{\mathbf{k}} &=&\varepsilon _{\mathbf{k}}-\mu +2\mathcal{T}_{4}(\mathbf{k}%
)N_{0}+3\mathcal{T}_{6}(\mathbf{k})N_{0}^{2}+...  \label{Ak} \\
\mathcal{T}_{4}(\mathbf{k}) &=&\Phi _{4}(\mathbf{k},\mathbf{0};\mathbf{k},%
\mathbf{0}),  \nonumber \\
\mathcal{T}_{6}(\mathbf{k}) &=&\Phi _{6}(\mathbf{k},\mathbf{0},\mathbf{0};%
\mathbf{k},\mathbf{0},\mathbf{0})...  \nonumber
\end{eqnarray}%
and
\begin{eqnarray}
B_{\mathbf{k}} &=&\mathcal{S}_{4}(\mathbf{k})N_{0}+\mathcal{S}_{6}(\mathbf{k}%
)N_{0}^{2}+...  \label{Bk} \\
\mathcal{S}_{4}(\mathbf{k}) &=&\Phi _{4}(\mathbf{k},\mathbf{-k};\mathbf{0},%
\mathbf{0}),  \nonumber \\
\mathcal{S}_{6}(\mathbf{k}) &=&\Phi _{6}(\mathbf{k},\mathbf{-k},\mathbf{0};%
\mathbf{0},\mathbf{0},\mathbf{0})...  \nonumber
\end{eqnarray}%
Diagonalizing the quadratic form (\ref{QuadForm}) by the linear canonical
transformation $b_{\mathbf{k}}=u_{\mathbf{k}}c_{\mathbf{k}}+v_{\mathbf{k}}c_{%
\mathbf{-k}}^{\dagger }$ , we find that
\begin{equation}
\mathcal{H}_{2}=U_{2}+\sum\limits_{\mathbf{k\neq 0}}\widetilde{\varepsilon }%
_{\mathbf{k}}d_{\mathbf{k}}^{\dagger }d_{\mathbf{k}},  \label{newquadratic}
\end{equation}%
where
\begin{equation}
U_{2}=\frac{1}{2}\sum\limits_{\mathbf{k\neq 0}}\left( \widetilde{\varepsilon
}_{\mathbf{k}}-A_{\mathbf{k}}\right)  \label{U0}
\end{equation}%
and
\begin{equation}
\widetilde{\varepsilon }_{\mathbf{k}}=\mathrm{sign}(B_{\mathbf{k}})\left( A_{%
\mathbf{k}}^{2}-B_{\mathbf{k}}^{2}\right) ^{1/2}
\label{quasiparticle spectrum}
\end{equation}%
is the spectrum of quasiparticles. From this formula obviously follows the
following criterion of the condensate stability:
\begin{equation}
B_{\mathbf{k}}>0,\text{ \ }A_{\mathbf{k}}^{2}-B_{\mathbf{k}}^{2}>0\text{\ }.
\label{stability}
\end{equation}%
For the theory with four-magnon interactions we obtain
\begin{equation}
\mathcal{S}_{4}(\mathbf{k})>0,\text{ \ }\left[ 2\mathcal{T}_{4}(\mathbf{k})%
\right] ^{2}-\text{\ }\left[ \mathcal{S}_{4}(\mathbf{k})\right] ^{2}>0.
\label{fourmagnon criterion}
\end{equation}%
If this condition does not work (for an attractive interaction, $\mathcal{S}%
_{4}(\mathbf{k})<0$), the six-magnon interactions which are repulsive in
ferro- and antiferromagnets (due to specificity of Holstein-Primakoff
representation of spin operators by the bose operators) can make the system
stable:
\begin{eqnarray}
\mathcal{S}_{4}(\mathbf{k})+\mathcal{S}_{6}(\mathbf{k})N &>&0,\text{ \ \ }
\label{criterion 6-1} \\
\left[ 2\mathcal{T}_{4}(\mathbf{k})+3\mathcal{T}_{6}(\mathbf{k})N\right]
^{2}-\left[ \mathcal{S}_{4}(\mathbf{k})+\mathcal{S}_{6}(\mathbf{k})N\right]
^{2} &>&0.  \nonumber
\end{eqnarray}%
In this case one can expect a sharp appearence of the uniform condensate at $%
N>-\mathcal{S}_{4}(\mathbf{k})/\mathcal{S}_{6}(\mathbf{k})$. A detailed
analysis will be published elsewhere.

\section{Discussion}

The Bose-Einstein condensate of magnons with $ k = 0$  is a uniform precession of the magnetic moment in the effective magnetic field.
 How does this precession differ from the usual precession of the magnetic moment? 
It is known that the uniform precession of a magnetic moment deviated from equilibrium precesses in an effective field 
and is gradually damped due to losses in the magnetic material and radiation damping. 
In this case we have an excited coherent state of magnons with $k = 0$, and the magnons of the entire 
spectrum are in thermodynamic equilibrium with the chemical potential $\mu = 0$.
In the case of BEC, we also have an excited coherent state of magnons with $k=0$, but this state arose as a result 
of a change in the density of magnons and their chemical potential becomes $\mu=\varepsilon_0$.
In this case, the losses of the precessing condensate in the magnetic material disappears, 
the spin superfluidity arises. There remains only a weak radiation damping \cite{KalafatiSafonov, KSjmmm}, 
which leads to a long-lived coherent precession. 
This long-lived state was first observed experimentally in antiferromagnetic $^3$He-B \cite{HPD}.

In this paper we have focused on critical conditions of BEC in ferro- and
antiferromagnets, bulk and thin films. Recent similar analysis \cite{critSN}
demonstrated a good agreement with experiments for nuclear magnon BEC. Let
us list main results.

1. We have shown that in general the only highly populated region of the
spectrum (which has been previously considered for particular cases in Refs.%
\cite{review5},\cite{Kopec}) is responsible for the formation of any BEC.
The high population approximation, $n_{\mathbf{k}}\simeq k_{B}T/(\varepsilon
_{\mathbf{k}}-\mu )$, however, does not mean that we deal with the classical
physics, this is an expression for the Bose-Einstein distribution if $n_{%
\mathbf{k}}\gtrsim 1$. The opposition of high density of particles or
quasiparticles to their high population makes the classical criterion of
Maxwell-Boltzmann statistics inapplicable which means that the Bose-Einstein
condensation condition $\mu =\min \varepsilon _{\mathbf{k}}$ always occurs
in the degenerate bose gas. The high population approximation substantially
simplifies analysis of the critical conditions and can simplify magnetic
dynamics simulations of systems with BEC by the use of classical variables
instead of operators. So far as the BEC occurs in the $k$-space, the most
convenient form of simulation is the use of kinetic equations for $n_{%
\mathbf{k}}$ with the integrals of magnon-magnon collisions (see, e.g., (\ref%
{INT st})), sources of pumping and relaxation \cite{KalafatiSafonov},\cite%
{YIGBEC2}.

2. We have found that the condition of magnon BEC in one of the most
interesting materials, perpendicular magnetized YIG thin film, is fulfilled
at a small angle, when signals are usually treated as excited spin waves
\cite{Fetisov}.

3. We have estimated that in hematite, high-temperature antiferromagnet, the
BEC should occur at much lower level of the magnon excitation compared to
that of ferromagnetic YIG. We believe that this theoretical prediction can
open a new direction of purposeful studies for fundamental and applied
research.

4. We have generalized the Bogoliubov's BEC theory to the case of
multi-particle interactions. According to this theory the uniform
Bose-Einstein condensate is unstable if the four-particle interaction is
attractive. This situation sometimes takes place in ferro- and
antiferromagnets. The account of six-magnon interactions (which are
repulsive in magnets) can resolve the problem of BEC stability. In
principle, there are also factors of time, the sample size, and relaxation
for the BEC instability has to be developed in the system.

One more fundamental problem that remains in this field is to connect
microscopic and phenomenological description of bose system with
Bose-Einstein condensate. This problem exists for a long time and it's
solution, say, on the base of magnon systems, will advance the understanding
and progress in all areas where the BEC is manifested. The solution of this
problem will help understand the spin superfluidity as a property of the
magnetic system that accompanies BEC of quasi-equilibrium magnons.

In conclusion, we emphasize that the Bose-Einstein condensation of
quasi-equilibrium magnons is a fundamental law of physics. BEC appears due
to quantum statistics of quasiparticles in magneto-ordered systems and can
exist at room, or even higher temperatures. One of the most intriguing
properties of the BEC is a superfluid spin current, a coherent quantum flow
of energy and information. This understanding of magnon BEC in different
magnetic materials can be very useful for spin transport and magnonic
quantum devices. An interest in the spin currents, the magnetization
projection transfer in magnetic materials, is growing every year.

\acknowledgments The authors wish to thank Grigory Volovik 
and  anonimous reviewer for helpful comments.
 For Yu. M. B. this work was financially supported by the Russian
Science Foundation (grant RSF 16-12-10359).

\end{document}